\documentclass[a4paper,fleqn]{cas-sc}
\usepackage[super]{natbib}
\newcommand{\etal}{\textit{et al.}}
\def\tsc#1{\csdef{#1}{\textsc{\lowercase{#1}}\xspace}}
\tsc{WGM}
\tsc{QE}
\begin{document}

\let\WriteBookmarks\relax
\def\floatpagepagefraction{1}
\def\textpagefraction{.001}

\shorttitle{}    

\shortauthors{Akshat Dubey \etal}

\title [mode = title]{A Nested Model for AI Design and Validation}  


% \tnotetext[1]{Correspondence: DubeyA@rki.de (A.D.)} 
\author[1,2,3]{Akshat Dubey}[orcid=0009-0008-4823-9375]
\ead{DubeyA@rki.de}

\author[1]{Zewen Yang}[orcid=0000-0001-9974-3231]
\ead{YangZ@rki.de}

\author[1,2]{Georges Hattab}[orcid=0000-0003-4168-8254]
\ead{HattabG@rki.de}

\affiliation[1]{organization={Center for Artificial Intelligence in Public Health Research (ZKI-PH) at Robert Koch Institute},
            addressline={Nordufer 20}, 
            city={Berlin},
         citysep={}, % Uncomment if no comma needed between city and postcode
            postcode={13353}, 
            country={Germany}}

\affiliation[2]{organization={Department of Mathematics and Computer Science, Freie Universität Berlin},
            addressline={Arnimallee 14}, 
            city={Berlin},
         citysep={}, % Uncomment if no comma needed between city and postcode
            postcode={14195}, 
            country={Germany}}
\affiliation[3]{organisation = {Lead author}}


% \author[1,2]{Akshat Dubey\fnref{firstfoot}}[orcid=0009-0008-4823-9375]
% \ead{DubeyA@rki.de}%[<options>]
% % Corresponding author indication
% % \cormark[1]

% % Footnote of the first author
% % \fnmark[1]{Lead conract}

% % Email id of the first author
% \ead{DubeyA@rki.de}
% \fntext[label5]{Lead author}

% % URL of the first author
% % \ead[url]{}

% % Credit authorship
% % \credit{Conceptualization of this study, Methodology, Software}
% % \credit{}
% \author[1]{Zewen Yang}[orcid=0000-0001-9974-3231]
% \ead{YangZ@rki.de}

% \author[1,2]{Georges Hattab}[orcid=0000-0003-4168-8254]
% \ead{HattabG@rki.de}

% % Footnote of the second author
% % Email id of the second author
% % URL of the second author

% % Credit authorship


% % Corresponding author text
% % Footnote text
\begin{abstract}
The growing AI field faces trust, transparency, fairness, and discrimination challenges. Despite
the need for new regulations, there is a mismatch between regulatory science and AI, preventing a
consistent framework. A five-layer nested model for AI design and validation aims to address these
issues and streamline AI application design and validation, improving fairness, trust, and AI adoption.
This model aligns with regulations, addresses AI practitioners’ daily challenges, and offers prescriptive guidance for determining appropriate evaluation approaches by identifying unique validity threats. 
We have three recommendations motivated by this model: authors should distinguish between layers when claiming contributions to clarify the specific areas in which the contribution is made and to avoid confusion, authors should explicitly state upstream assumptions to ensure that the context and limitations of their AI system are clearly understood, AI venues should promote thorough testing and validation of AI systems and their compliance with regulatory requirements.

\end{abstract}



% Keywords
% Main text
\maketitle

\section*{Introduction}\label{}
While artificial intelligence (AI) has grown tremendously in recent years, it has yet to reach its true potential in real-world use cases. This is due in part to a lack of trust and transparency, as well as fairness and fear of discrimination~\cite{bedue2022can}. 
AI has unique strengths and weaknesses, so there will always be a need to develop new regulations and change old ones. But with the benefits of AI come significant ethical and legal risks. 
As a result, there is an urgent need to address not only the regulatory policies that will facilitate the implementation of AI in real-world use cases but also how practitioners design and validate AI applications and workflows.
Several regulatory bodies from different countries are stepping in to establish a set of regulations for the implementation of AI in real-world applications. Researchers have shown an increased interest in unifying regulatory science and AI.

Surveys on the subject of AI and human-computer interaction have already been conducted. They draw attention to the drawbacks and difficulties with interactive machine learning and transparency in AI~\cite{weller2019transparency, bilgic2005explaining, bellotti2001intelligibility}. Many of them offer a set of conceptual and design guidelines to help ensure that intelligent systems are understandable and that human users are held accountable. Using a framework, Mohseni \etal. published an intriguing work that surveys and identifies the state of research on the junction of XAI and HCI~\cite{mohseni2021multidisciplinary}.To bring the iterative design and evaluation cycles in diverse Explainable AI teams to a close, they created a framework with detailed design rules along with assessment techniques.

AI systems are complex in nature and often involve multiple stakeholders. One of the key tools for managing complex systems is modularity. By distinguishing between activities that require extensive analysis and those that do not, modularity seeks to minimize the number of interdependencies that need to be examined. A specific type of modularity known as ``layering" involves the arrangement of different system components into parallel hierarchies. Most research on nested models is based on the layering approach~\cite{yoo2013protocol}.

To address the complexity of AI systems, the authors break down governance issues into smaller, more manageable parts, encourage shared accountability among stakeholders, and provide a framework for the creation of laws, policies, standards, and other guidelines that can be used in concert to guide the responsible development and application of AI technologies~\cite{gasser2017layered}. 
The model consists of three layers, namely social \& legal, ethical, and technical (including algorithms and data), which can be developed independently.  
They propose a modular approach that divides AI governance into social, legal, ethical, and technical layers to minimize risks and maximize benefits, making it more efficient and tractable.
The social and legal layer regulates AI by establishing institutions and norms within a legal framework. 
The Ethical Layer addresses ethical concerns for AI systems, ensuring fairness, accountability, and transparency. 
The Technical Layer, based on algorithms and data, promotes fairness and safeguards against discrimination. 
These layers work together to manage the societal impacts of AI and ensure its trustworthiness, accountability, and auditability.
However, we argue that the technical layer can't be addressed without addressing regulatory requirements. 
For example, one of the most common requirements set by AI regulators is AI transparency. While addressing the three layers independently, the technical layer may or may not address the challenge of transparency if the technical layer is addressed first, leading to high uncertainty in development.

Recent research from Wang\etal presents a user-centric XAI as a nested model. 
The study is limited to drug repurposing using graph neural network (GNN)~\cite{corso2024graph}. Although the authors address the domain context, which is very important for XAI, they don't address the regulations, which is another important perspective~\cite{wang2022extending}.

Several research efforts from the fields of Human-Computer Interaction (HCI) and Explainable AI (XAI) have highlighted current bottlenecks in involving humans in understanding, decision-making, validation, etc. or having humans in the loop for AI applications and workflows.
There is an influential work by Liao~\etal, on the intersection of HCI and XAI in the form of XAI-Question Bank (XAI-QB). XAI-QB aims to solve the challenge of achieving full explicability through algorithm-informed prototypical questions~\cite{xaiqb_liao}.
XAI-QB was further advanced and was extended to include the prototypical questions for the end-users which they may come across while interacting through an interface~\cite{sipos2023identifying}.
While the need for guidelines and regulatory frameworks has been addressed by Lennerz~\etal, one of the problems is how AI is implemented and how results are communicated, which further hinders AI adoption~\cite{lennerz2022unifying}.
The authors specifically mention that regulatory concepts are necessary for AI researchers, as these concepts allow risk and safety concerns to be addressed and understood by the regulations proposed by the US and Europe. 

At the moment, however, the fields of regulatory science and AI are diverging, with no major overlap in sight.
A growing trend has been observed in discussions that express an urgent need for work at the intersection of regulatory science and AI~\cite{reg_1, reg_2, reg_3, reg_5, reg_7, reg_8, reg_10, reg_22}. 
When it comes to regulating AI, many regulatory bodies are stepping in and making it mandatory to comply with laws that require explanations or interpretations to be given to users when confronted with algorithmic output~\cite{broniatowski2021psychological}. 
The intersection of artificial intelligence (AI) and regulation requires massive work more than ever due to the emergence of new AI laws~\cite{reg_13, reg_16, reg_11, reg_21}. There is a growing trend of discussions that express an urgent need for the aforementioned intersection~\cite{reg_17, reg_18, reg_19}. 
In response to the growing need for updated regulations to address ethical and legal risks that face obstacles due to a misalignment between regulatory science and AI, we have presented a structured, five-layer, nested model for the design and validation of AI that serves as a systematic guide for the assessment and validation of AI applications and workflows. 
It facilitates the identification of appropriate evaluation methodologies by identifying unique threats within each layer, thereby mitigating the inherent tensions between technological innovation and regulatory imperatives. 
In addition, the proposed model addresses concerns about fairness, trust, and the alignment of AI models with existing regulations. 
To address the challenge of bridging the gap between AI practitioners and regulators, two preliminary case studies include researchers who aim to develop reliable, wise, and trustworthy human-centered AI through ethical and theoretical guidelines with management strategies or software engineering practices.
These include audit trails to enable analysis of failures, software engineering workflows, verification and validation testing, bias testing to enhance fairness, and even explainable user interfaces~\cite{chakraborty2021bias, software_practices_bridging, article_bridging_ethical_ai}.

The proposed model for AI design and validation is the first of its kind, inspired by XAI-QB, which we extend to include the prototypical questions from a regulatory and domain perspective. 
In parallel with the regulations, it addresses the issues that AI practitioners face on a daily basis. 
The nested model consists of five layers, namely the regulation, the domain, the data, the model, and the prediction.
We have expanded regulatory and domain questions from the XAI-QB. 
In addition, we have grouped the prototypical questions from the XAI-QB based on the layer of the nested model of AI design and validation that they need to address.
The five-layer nested model for AI design and validation, including but not limited to its regulatory and domain-aware layers and questions, are our major contributions. 
The nested model for AI provides much-needed overlap and grounding, facilitating the design and validation of AI applications and workflows, and increasing fairness, trust, and AI adoption.

\section*{Regulation of AI}
Calls for appropriate regulation of AI have grown as awareness of its risks has increased. 
This regulation aims to ensure that AI is legal, ethical, and robust while minimizing potential harm and increasing legal certainty. 
Efforts are underway to establish global regulatory standards for AI, potentially leading to harmonization. 
Collaboration between government, industry, and civil society is essential for the responsible use of AI. 
Regulation is needed to protect consumers and society, provide a reliable framework for businesses, and understand the ethical and societal implications of AI. 
The complexity and risks associated with AI underscore the urgency of establishing best practices and a comprehensive framework for AI regulation that takes into account ethical, legal, and societal impacts~\cite{Smuha2019FromA,reg_6}.

The EU has taken a leading role in developing AI regulations, with a diverse regulatory landscape and a focus on protecting fundamental rights. 
The General Data Protection Regulation (GDPR) includes provisions that address the legitimacy of algorithmic decision-making, emphasizing the right to human intervention and meaningful information for individuals. 
The European Commission's Ethics Guidelines for Trustworthy AI and the proposed AI Act aim to ensure the responsible and transparent use of AI, introducing principles such as human agency, technical robustness, privacy, transparency, and accountability. 
These efforts aim to create a framework for trustworthy AI that benefits society and the environment~\cite{regulation2016regulation, nannini2023explainability,smuha2019eu,act2021proposal}.

In the United States, AI regulation has relied primarily on self-regulation by industry stakeholders, leading to criticism for a lack of rigorous regulatory oversight. The Defense Advanced Research Projects Agency (DARPA) has initiated the first research program in XAI to address the challenge of opaque yet effective AI systems, with the goal of developing machine learning techniques for more explainable models. The National Institute of Standards and Technology (NIST) has also emphasized the importance of explainability, proposing principles to ensure that AI systems provide understandable explanations for their results. While the US regulatory landscape is focused on promoting innovation, the EU's emphasis on XAI and ongoing policy discussions in the US and UK reflect a growing recognition of the importance of explainability in AI systems~\cite{nannini2023explainability, heer2018partnership,gunning2019darpa,phillips2020four}.

\section*{Nested Model for AI Design and Validation}
\subsection*{Motivation}
\label{motivation}
The EU is developing AI regulations with key requirements that align with the principles of Explainable AI (XAI), but addressing both ethical and technical requirements may require a focus on XAI and human-computer interaction (HCI) to ensure that AI upholds broader values such as accountability, human rights, and sustainable innovation.
At the intersection of XAI and HCI, the XAI-QB serves as a valuable tool for understanding user requirements for XAI.
It provides a set of algorithm-informed questions to achieve user-centered explainability in AI applications. 
To address the perspective of domain and regulatory authorities, and inspired by both Munzner's nested model for visualization and the XAI-QB, we propose a five-layer nested model for the design and validation of AI applications and workflows.
\subsection*{Our Work}
\label{sec:ethical_technical}
Our work has taken full advantage of the state of the art and the effectiveness of the nested model.
We extended the XAI-QB to include regulatory and domain-layer questions, aiming to help organizations address the needs of regulators when implementing AI. 
This user-centric approach provides a hierarchical guide for stakeholders to elicit end-user and regulatory needs, while also highlighting technical barriers and emphasizing the importance of a human-centered approach to regulation using XAI. 
The nested model supports specifying requirements for building AI applications and identifies opportunities for collaboration between the HCI and AI communities, industry practitioners, and academics to advance the field of AI.
\begin{figure}
\centering
\includegraphics[]{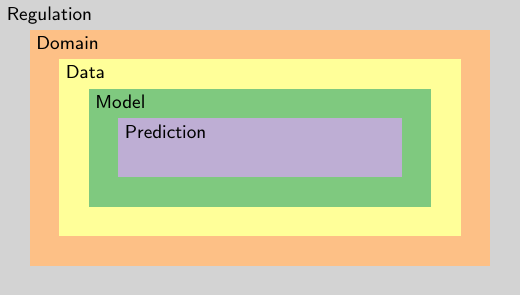}
\caption{Five-layer nested model for AI design and validation.}
\label{fig:nested_model}
\end{figure}
%One of the key points for this model to be successful in helping users meet their needs is to use the methodologies that answer potential research questions. There may be multiple potential threats that can lead to non-compliance. 
Our nested model (Refer figure:~\ref{fig:nested_model}) is divided into five distinct layers.
The model can be accompanied by questions. 
These questions should be answered at each layer.
Although these questions can be modified or extended depending on the use case and user-specific requirements, they help to satisfy regulatory requirements.

The regulation layer is responsible for making the AI workflow compliant. 
At this layer, we categorize the rules into ethical and technical regulations. 
Ethical rules should be specified and addressed at this layer. 
This layer prohibits access to subsequent layers if the ethical rules are not addressed. 
Once the ethical rules have been addressed with the appropriate infrastructure and various methodologies, the user should proceed to subsequent layers to address the technical rules.
\begin{enumerate}
\item Which regulation should be specified for this AI workflow?
\item Which country has specified this regulation?
\item Can you list out the key requirements set up by the regulation?
\item Can you categorize the guidelines into ethical and technical ones?
\item Who are the stakeholders involved in this AI workflow?
\end{enumerate}

The domain layer ensures that any process taking place in the nested model is within the scope of that domain. 
This nested model will allow for explainability at the domain layer. Different domains, in different settings, may have different answers to the XAI questions. At this layer, the domain expert is solely responsible for achieving certain goals related to the domain layer. Some of the goals may be to list the needs and requirements. After that, the domain expert should decide on the appropriate model validation metrics.
\begin{enumerate}
\item What are the specific requirements within the domain?
\item Does the domain encompass high-stakes areas such as healthcare or finance?
\item What are the potential risks associated with the domain?
\item Are there any pre-existing assumptions that are necessary?
\item Is feedback from domain experts a requirement for this process?
\end{enumerate}

The data layer aims to explain the data by summarizing and analyzing the data and providing insights into the data. 
This layer helps the user understand the data, the biases involved, how to mitigate the biases, the distribution, the limitations of the data, and what domain knowledge is contained in the data. This layer involves both the ML practitioner and the domain expert. The domain expert should list the limitations of the data, and both the ML practitioner and the domain expert should decide whether the data can be used for the specific application, The domain expert should provide the prior domain knowledge in the form of knowledge graphs to the ML practitioner, who will then incorporate this domain knowledge during training of the ML model. The data should be represented visually in the simplest way possible by the ML practitioner, so that the domain expert has a clear understanding of the data, which may help them to list the limitations or drawbacks of the data or the bias in the data, the ML practitioner will then mitigate the bias. One of the easy ways to detect and mitigate the bias in the data is to use the capabilities of IBM's AI Fairness 360 (AIF360) library~\cite{aif360-oct-2018}. This library can be used to remove the bias from both the data and the model. The data layer has the associated XAI questions.

\begin{enumerate}
\item What type of information is contained within the data?
\item What inferences can be drawn from this data?
\item Which aspects of the data are the most significant?
\item How is the information distributed within the data?
\item Is it feasible to enhance the model's performance by reducing the number of dimensions?
\item Could the use of data summarization techniques provide a more effective explanation?
\end{enumerate}
    
The model layer aims to explain the inner workings of the model, the parameters involved and their meaning, the interpretability at the model layer, and which model maintains the balance between performance and interpretability. 
This layer involves an ML practitioner. One of the most common baselines to get started with model-layer explainability might be to answer whether the interpretable models can be used instead of black-box models~\cite{10.5555/3635637.3663066}. If the interpretable model can be used, then the hyper-parameter analysis should be done thoroughly to get the best result. There is often a trade-off between interpretability and performance. If performance is the most important goal at this layer, then post hoc methods can be used to achieve interpretability; otherwise, the interpretable models should undergo hyper-parameter tuning to get the best results without black box models.
\begin{enumerate}
\item What attributes render a parameter, objective, or action important to the system?
\item At what point did the system assess a parameter, objective, or action, and when was it disregarded by the model?
\item What are the repercussions of altering a decision or modifying a parameter?
\item How was a specific action executed by the system?
\item How are these model parameters, objectives, or actions interconnected?
\item What elements does the system consider (or exclude) when making a decision?
\item What methods does the system employ or avoid to accomplish a goal or inference?
\end{enumerate}

The prediction layer aims to explain the reason for a particular prediction, how certain inputs affect the prediction, whether the reason is sufficient for the conclusion or decision, what variables are involved behind the prediction, and how the prediction changes under certain considerations or criteria. 
At this layer, the domain expert relies on the ML practitioner for results and a deeper understanding of the predictions. This layer would help answer the prediction-layer questions.
\begin{enumerate}
\item What factors contribute to the importance of a parameter, objective, or action within the system?
\item When was a parameter, objective, or action evaluated by the system, and when was it rejected by the model?
\item What are the consequences of changing a decision or adjusting a parameter?
\item How was a specific action carried out by the system?
\item How are the parameters, objectives, or actions within the model interconnected?
\item What elements does the system take into account (or disregard) when making a decision?
\item What methods does the system utilize or avoid to achieve a specific goal or inference?
\end{enumerate}
\subsection*{Addressing the question "How the nested model solve the problem?"}
\begin{enumerate}
    \item Once the regulations are followed or the key requirements have been introduced, only then can the next layer, i.e., the domain layer, be entered.
    \item The nested model prohibits entering further sub-layers until the goal of the previous layers has been achieved. For example, after accomplishing the specific goal of the domain layer, which is to define applications and requirements, the user can proceed to the data layer to accomplish certain goals within the scope of the data layer.
    \item The nested model is built on the foundation of the XAI-QB, which allows the user to answer specific questions that lead to a clear understanding of the rules, goals, needs, requirements, solutions, and conclusions.
    \item nested model conforms to regulatory guidelines. This leads to solving certain challenges that arise when using AI workflows. This enables AI to be compliant, trustworthy, accountable, non-discriminatory, appropriate for human equality, robust \& secure, and transparent AI that operates under human agency \& oversight.
    \item The needs of regulators are addressed in the form of prototypical questions. It uses the human-computer interaction approach to eliminate the threat posed by the lack of human agency and oversight. 
    \item The nested model can be used as a guidance tool to support the need for specification work to create AI applications that meet the key requirements of regulatory authorities.
    \item The nested model addresses "transparency and explainability" from both the regulatory and AI perspectives.    
    \item The nested model is a prescriptive guideline that bridges the gap between AI regulations and AI.
\end{enumerate}
\subsection*{Outcomes of the Nested Model for AI}
The outcomes of the Nested Model are as follows:
\begin{itemize}
    \item It helps define a common baseline for the adoption of AI in real-life use cases.
    \item It acts as a common intersection between regulatory authorities (and their regulations), XAI, and AI practitioners by bringing these divergent fields together.
    \item Reduces the chances of AI implosion by addressing the issues and potential threats with each layer. Mitigates the problems of non-transparency, unfairness, and discrimination.
    \item Evaluates the AI workflow, not only through evaluation metrics but also through key requirements of the regulations.
\end{itemize}
\section*{Implementation}
To implement the nested model for the design and validation of AI workflows for AI governance we need to define regulations into ethical and technical.
Modern regulations are a combination of both ethical and technical implementations. Ethical regulations and technical regulations in the context of AI can be distinguished based on their focus and objectives. Ethical regulations are often concerned with guiding the moral principles and values associated with AI development and deployment, while technical regulations focus on specific technical aspects and requirements to ensure the responsible and safe use of AI systems~\cite{ethics_technical}.     
\begin{itemize}
    \item Ethical Requirements:
        Focus: Ethical regulations are primarily concerned with promoting values, principles, and norms associated with responsible AI development and usage. This may include considerations of fairness, transparency, accountability, privacy, and the broader societal impact of AI technologies.
        Objective: The primary goal is to ensure that AI systems align with ethical standards and do not cause harm or violate fundamental human rights. Ethical guidelines provide a framework for developers and organizations to make morally sound decisions throughout the AI life cycle. The ethical key requirements could be solved with methodologies listed by Al Alhamed \etal~\cite{article_bridging_ethical_ai}.

    \item Technical Requirements:
        Focus: Technical regulations, on the other hand, are more specific and detail-oriented. They focus on the technical aspects of AI systems, such as algorithms, data quality, safety measures, and other technical requirements.
        Objective: The objective of technical regulations is to set standards and requirements that AI developers must follow to ensure the robustness, security, and reliability of AI systems. These regulations are often designed to prevent technical issues, biases, and potential risks associated with AI deployment.
\end{itemize}
Using the above discussion as a reference, we define steps to implement the nested model, referencing questions at each layer as needed.
Steps to follow:
\begin{enumerate}
    \item Define the regulations and key requirements at the layer of the regulation.
    \item Categorize the key requirements into ethical and technical requirements.
    \item At the layer of the regulation, address the ethical requirements first.
    \item After proceeding into the domain layer, the domain expert will list down the domain-specific requirements, which will act as a reference for AI practitioners.
    \item Map all the technical key requirements from the regulations to the sub-layer, namely, data, domain, and prediction (Refer figure:~\ref{fig:mapping_technical_layers}).
    \item Address the specified technical key requirements at each sub-layer using appropriate methodologies.
\end{enumerate}
\subsection*{Use Case: Europe Union Requirements for Trustworthy AI}
\begin{figure}
    \centering
    \includegraphics[]{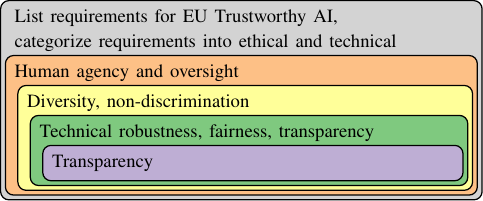}
    \caption{Mapping of technical requirements to the layers of the nested model for AI design and validation.}
    \label{fig:mapping_technical_layers}
\end{figure}

Listing out the steps:
\begin{enumerate}
    \item Define the regulations and key requirements at the layer of the regulation :
    Based on fundamental rights and ethical principles, the Guidelines list seven key requirements that AI systems should meet to be trustworthy:
    \begin{itemize}
        \item Human agency and oversight
        \item Technical robustness and safety
        \item Privacy and data governance
        \item Transparency 
        \item Diversity, non-discrimination, and fairness
        \item Societal and environmental well-being
        \item Accountability
    \end{itemize}
    \item Categorize the key requirements into ethical and technical requirements:
    After the ethical and technical requirements are satisfied, they will result in accountability.
    \begin{itemize}
        \item Ethical requirements: Privacy and data governance, societal and environmental well-being, safety.
        \item Technical requirements: Human agency and oversight, technical robustness, transparency, diversity, non-discrimination, fairness
    \end{itemize}
    \item Address the ethical requirements.
    \item Domain experts list the domain-specific requirements.
    \item Map the technical requirements with the sub-layers namely Data, Model, and Prediction (Refer figure:~\ref{fig:mapping_technical_layers}).
    \item Address the specific technical requirements at the specific sub-layer using appropriate methodologies. When all the key requirements (ethical and technical) are achieved, it results in accountability.
\end{enumerate}

\subsection*{Potential Threats}

\subsubsection*{Regulation layer}
\begin{figure}
    \centering
    \includegraphics[]{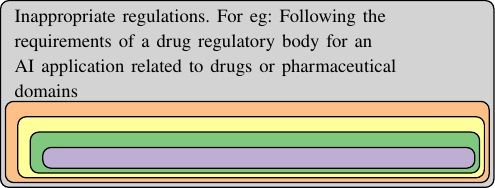}
    \caption{Potential threats at the regulation layer.}
    \label{fig:regulation_threats}
\end{figure}

The regulation layer helps to ensure compliance with the AI workflow. 
Regulatory threats need to be carefully considered (Refer figure:~\ref{fig:regulation_threats}).
As operational burdens and regulatory requirements increase, the AI infrastructure will need to also adapt; including data management, privacy, security, and transparency standards.
To ensure compliance and foster an environment conducive to evaluating, validating and ultimately adopting AI technologies, organizations must be prepared to address these issues collectively. The AI application must align with the expectations of both the AI regulatory body and the domain regulatory body for which the AI is being developed.
\subsubsection*{Domain layer}
\begin{figure}
    \centering
    \includegraphics[]{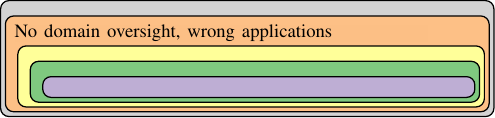}
    \caption{Potential threats at the domain layer.}
    \label{fig:domain_threats}
\end{figure}

The domain layer sets boundaries on processes that fall within the domain's purview (Refer figure:~\ref{fig:domain_threats}).   

\subsubsection*{Data layer}
\begin{figure}
    \centering
    \includegraphics[]{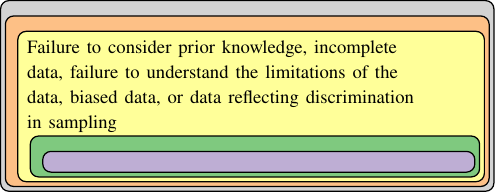}
    \caption{Potential threats at the data layer.}  
    \label{fig:data_threats}  
\end{figure}

The data layer strives to explain the data by summarizing and analyzing the data and providing insights into the data (Refer figure:~\ref{fig:data_threats}). 

\subsubsection*{Model layer}
\begin{figure}
    \centering
    \includegraphics[]{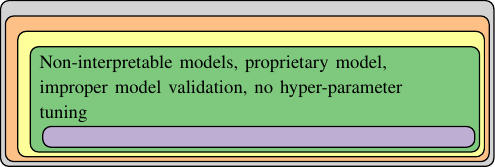}
    \caption{Potential threats at the model layer.}   
    \label{fig:model_threats}
\end{figure}

The model layer seeks to explain the inner workings of the model, the parameters involved and their meaning, the interpretability of the model, and whether the model maintains the balance between performance and interpretability (Refer figure:~\ref{fig:model_threats}).

\subsubsection*{Prediction layer}
\begin{figure}
    \centering
    \includegraphics[]{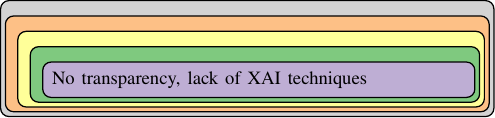}
    \caption{Potential threats at the prediction layer.}  
    \label{fig:prediction_threats}
\end{figure}

The prediction layer aims to explain the reason for a particular prediction, how certain inputs affect the prediction, whether the reason is sufficient for the conclusion or decision, what variables are involved behind the prediction, and how the prediction changes under certain considerations or criteria (Refer figure:~\ref{fig:prediction_threats}).
    
\section*{Examples}
We now analyze several previous use cases in terms of our
model, to provide a concrete example.
\subsection*{AI Model by Google LLC to detect retinopathy.}
\begin{figure}
    \centering
    \includegraphics[]{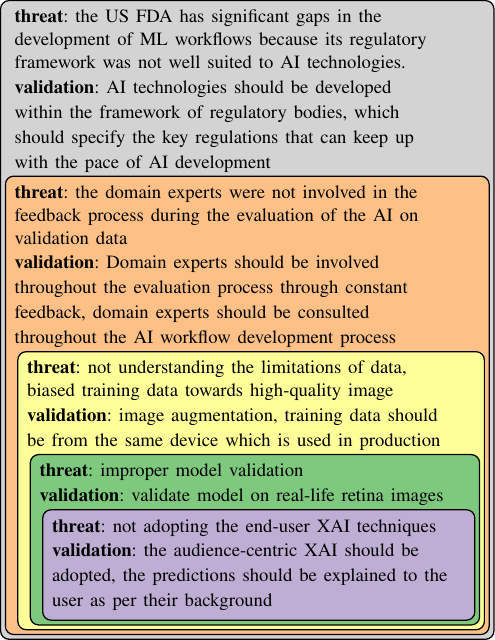}
    \caption{Potential threats: Google AI model to detect diabetic retinopathy.}   
    \label{fig:example_google}
\end{figure}

Diabetic retinopathy is a disease of the retina caused by diabetes that leads to vision impairment or loss. 
During the development of the workflow, the ethical and technical requirements were neither listed nor addressed.
Google's deep learning model for the detection of diabetic retinopathy failed for several reasons. The model had been trained on high-quality, high-resolution eye scans, but in real-life clinics, the images captured by nurses differed in quality and lighting conditions, leading to a significant disparity between the training data and real-life data. 
The model was approved by the Food and Drug Administration (FDA) of the United States (Refer figure:~\ref{fig:example_google}). 
 
While the FDA is essential to the regulation of AI in healthcare, additional specialized oversight mechanisms are needed to address the unique issues presented by AI technologies. This includes ensuring that AI applications are clear, safe, and efficient. Adaptive AI and machine learning (ML) technologies do not lend themselves well to the FDA's traditional regulatory framework, as they can change over time in response to new data. This makes it difficult to ensure the effectiveness and safety of these dynamic systems. While there are still many unanswered questions, the FDA is currently addressing the regulatory issues raised by AI in healthcare. However, more comprehensive and transparent regulatory processes are required to handle the dynamic nature of AI as a technology~\cite{fda, problem_fda}

Additionally, the model was not validated on real-life data, which could have been verified by domain experts.
This discrepancy between training data and real-life conditions, as well as the lack of validation of real-life data, contributed to the model's failure to accurately detect diabetic retinopathy in a clinical setting~\cite{diabeticfailure}.
The integration of user-centric explainable AI (XAI) approaches and collaboration with domain experts can enhance the evaluation of AI systems. These approaches prioritize understanding user needs, identifying explainability needs, fostering collaboration between domain experts and AI researchers, capturing domain knowledge, identifying model inaccuracies, addressing the explanations, addressing biases, developing adaptive explanations, and creating a comprehensive XAI evaluation framework~\cite{roselli2019managing}. By involving domain experts in the design process, researchers can create more effective and understandable explanations, improve model accuracy and robustness, and address social biases in decision-making. XAI can also develop context-aware and adaptive explanations that match user mental models and expertise levels, ensuring relevance and understanding across different user groups. 
By leveraging these approaches, researchers and practitioners can create more transparent and trustworthy AI systems that better serve their intended users, eventually improving adoption in the target knowledge domain~\cite{parikh2019addressing}.

\subsection*{Zillow Group, Inc. House Price Forecasting Model}
\begin{figure}
    \centering
    \includegraphics[]{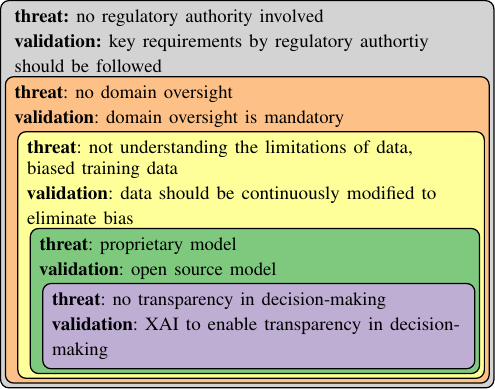}
    \caption{Potential threats: Zillow Group, Inc. ML model failed to forecast house prices.}  
    \label{fig:example_zillow}
\end{figure}

Zillow suffered a significant loss of over \$500 million in its home-flipping business due to the failure of its non-transparent proprietary forecasting algorithm~\cite{troncoso2023algorithm}.
The loss was due to the algorithm's inability to accurately predict home prices, resulting in overpayment for homes and financial volatility. 
This raised concerns about the reliability of AI models in critical business decisions, prompting a reevaluation of the use of AI in high-stakes operations and highlighting the importance of considering potential limitations and risks. 
The failure underscores the need for reliable and transparent AI algorithms and emphasizes thorough evaluation and risk assessment when integrating AI into decision-making processes. 
The event serves as a cautionary tale for organizations relying on AI for critical business strategies, highlighting the potential consequences of inadequate algorithm performance and the need to maintain a critical perspective on AI's capabilities and limitations (Refer figure:~\ref{fig:example_zillow}).
 
\subsection*{The Feature Cloud Platform for Federated Learning }
\label{federated_learning}
\begin{figure}
    \centering
    \includegraphics[]{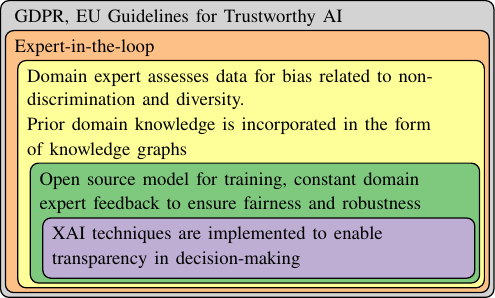}
    \caption{Validation of The Feature Cloud Platform through the Nested Model for AI design and validation.} 
    \label{fig:example_feature_cloud}  
\end{figure}

This platform takes advantage of federated learning and is an impressive work by Holzinger~\etal~\cite{feature_cloud} presents a comprehensive exploration of the integration of domain knowledge graphs into deep learning for improved interpretability and explainability using Graph Neural Networks (GNNs). 
Federated Learning (FL) protects privacy by transmitting only model updates, reducing the risk of data breaches. Its decentralized approach enhances data security and compliance. 
FL promotes collaboration between devices, improving generalization with diverse data sets. 
In addition, it improves the learning algorithm by incorporating explanations and conceptual knowledge for better interpretability~\cite{li2020review}.
The authors focus on using a protein-protein interaction (PPI) network to enrich deep neural networks for classification, enabling the detection of disease sub-networks using explainable AI.
This work addresses the potential threats at each layer of the nested model and validates them accordingly (Refer figure:~\ref{fig:example_feature_cloud}).
 
 \begin{enumerate}
     \item Regulations: GDPR and Europe Union Requirements for Trustworthy AI.
     \begin{itemize}
         \item Ethical Requirements: Privacy and data governance,
         societal and environmental well-being, safety.
         \item Technical Requirements: Human agency and oversight, technical robustness, transparency, diversity, non-discrimination, fairness
     \end{itemize}
     \item Addressing Ethical Requirements: Privacy, data governance, and safety are addressed through a federated learning approach. This use case accelerates biomedical research, which in turn benefits humanity and enables societal well-being. 
     In addition, by decentralizing the training process, federated learning has the potential to be more environmentally friendly, making it a promising approach for reducing the carbon footprint of machine learning model training~\cite{qiu2023first}.
     \item Addressing the technical requirements: The framework included an expert-in-the-loop approach to develop AI workflows under the supervision of a domain expert, allowing for human agency and oversight. The domain expert evaluates data for biases related to non-discrimination and diversity. 
     Knowledge graphs are implemented to incorporate prior domain knowledge. 
     The potential threat at the data layer is validated. At the model layer, models are evaluated for fairness and technical robustness through continuous feedback from the domain expert. 
     Open-source models are preferred for training. Finally, the XAI is integrated to make the AI process transparent and trustworthy.

In addition, this work also answers the extended XAI-QB in a summarized fashion. 
The questions answered are "What was not done and why?", "What problems were solved?", "What was difficult?", "What did we learn?", and "What are the limitations and future results?".
 \end{enumerate}
 After validating this work against our nested model, we concluded that this work presents an appropriate paper at the intersection of HCI and AI that meets the key requirements set by the EU Trustworthy AI Guidelines and GDPR.

\section*{Discussion}
We have discussed the potential threats and their validation methodologies, but we assume that the potential threats are a non-exhaustive list. 
There can be many potential threats and validation methodologies that may exist at the given layer, and these potential threats and validation methodologies should be mapped to one of the layers of the nested model for AI design and validation. 

Considering the need for the domain expert as an expert-in-the-loop, it is also possible that a single domain expert may not be able to list all the domain-specific requirements, model validation strategies, visualization techniques, etc., so we recommend multiple iterations of the nested model for AI design and validation. 
In addition, the XAI-QB should be kept in mind and the specific sub-type of the topic can be extended according to the requirements. 
For example, if a domain-specific data visualization is desired, then this question should be added to the XAI-QB in the data subtopic and then the appropriate methods should be used at the data level to address this question. 
The advantage of this framework is that it allows users to evaluate the AI workflow in human terms in collaboration with XAI-QB, which is an algorithmic question bank. 

In addition, there may be questions such as "Sometimes we don't know the problem with the data until we start training". 
One answer is that we are still at the data level because the issue with the data is still there in the training of the model. 
After the training of the model, it reflects that there is a problem with the data, then technically you are using the ML model to analyze the data for more potential threats, this will act as a methodology for the identification of unknown or more potential threats. 
Accordingly, a question should be added to the XAI-QB under the data subtopic that asks, "How do you identify unknown potential threats with the data that are not identifiable by normal bias identification methods?"
The possibility of unstated biases is an open question, and some may be implicit threats.

The process of navigating the nested model for AI enables validation.
We name this process as \textit{"eliminating the potential threats of one layer through the functionality of the next layer"}.
Successfully navigating through each layer of the nested model layer for AI means that one has successfully resolved the threats present at each layer.

Another important point is that the regulations don't clearly differentiate between the terms used for ethical and technical key requirements. 
For example, technical robustness could mean both robustness in terms of infrastructure and robustness in terms of AI or decision-making. So we assume that the requirements are specified from both technical and ethical perspectives.
Our nested model for AI design and validation is the only way to design the AI workflow and meet the listed key requirements according to the regulations. 
Our model addresses the regulations through XAI-QB and HCI, allowing AI to be evaluated in human terms.

While we have not listed the full list of algorithms or methods to address every possible potential threat, we argue that the nested model for AI is sufficient to warrant a rethink at each step of the conceptualization, design, and validation steps. 
The speed at which AI regulations will impact infrastructure, and the need to quickly comply with regulatory requirements, justifies the regulatory layer as an overarching gateway to validate AI systems and prepare them for the real world.
The threats to validity may vary depending on the choice of algorithms, and the potential threats may vary from use case to use case; our advice is to iterate the nested model for AI as often as necessary.

The nested model takes into account both the AI in production and the AI in development. 
AI regulations are designed to address both the development and production or deployment phases of AI systems. 
They aim to create a framework that promotes innovation while safeguarding against potential risks and ensuring ethical use of AI technology throughout its lifecycle. 
One of the examples that aligns with this scenario is the EU AI Act. 
It takes a comprehensive approach that covers the entire AI lifecycle. 
The legislation establishes rules for AI developers, deployers, and end-users. 
It also sets up governance structures at the national and international levels to oversee AI regulation. 
Furthermore, it creates mechanisms for ongoing assessment and adaptation of regulations as AI technology evolves. 
For instance, for AI in development, transparency is required in the AI model's design and training process. 
Additionally, the results and decision-making processes must be communicated effectively with the help of audience-centric XAI techniques.

There exist various regulations for the domains. 
One such example is pubmed regulations for the biomedical domain. 
Two seminal works focus on guidelines and best practices for health informatics and software development in healthcare, with a particular emphasis on artificial intelligence (AI) applications~\cite{hauschild2021fostering, hauschild2022guideline}.
These articles highlight the importance of creating reproducible and reusable biomedical software, fostering technology transfer in health informatics, and leveraging AI's potential in medical research and precision medicine.
Furthermore, the articles address the software life cycle in health informatics and the role of AI in supporting clinical decision-making, which can enhance patient monitoring, diagnostics, and prognostics. 
They also discuss crucial aspects often considered in healthcare software regulations, such as ensuring software reliability, maintaining data privacy and security, validating AI algorithms for clinical use, and implementing proper documentation and version control. 
It is crucial to acknowledge that the specific regulations pertaining to health informatics and AI in healthcare may vary by country and jurisdiction. 
Therefore, it is essential for developers and healthcare providers to consult with the relevant regulatory bodies to ensure that they are aware of the most up-to-date and applicable regulations when implementing health informatics solutions.

One could argue that questioning is not sufficient. However, formulating and asking the right questions brings up potential problems and assumptions from regulation all the way to predictions of an AI system. If a certain question or questions cannot be answered truthfully, or without reasonable doubt, the question(s) shall remain open to show transparency and potential pitfalls along every subsequent layer of the nested model for AI design and validation. Naturally, certain questions require a hard limit, so do not proceed to subsequent layers. This validation is meant to resolve any downstream problems when developing AI systems.

The future work would focus on developing an audience-centric XAI that takes into account the background of the end-users. 
The development of audience-centric XAI necessitates a comprehensive approach that prioritizes user needs and contexts. 
Key requirements include the implementation of user-centric design principles to tailor explanations for diverse user groups, the provision of real-time and actionable insights, and the utilization of scenario-based methods to elicit specific user requirements. 
The integration of human-computer interaction principles is of paramount importance for the creation of effective explanation interfaces. 
Furthermore, the assurance of compliance with legal and regulatory standards is of critical significance. 
By meeting these requirements, XAI systems can deliver customized, meaningful explanations that enhance user understanding and trust across various contexts, while also fulfilling legal obligations. 
This approach aims to bridge the gap between complex AI systems and end-users, making AI decisions more transparent and interpretable for different audiences.~\cite{maxwell2023meaningful}
It is not uncommon for there to be debate surrounding the provision of overly detailed explanations, which could potentially be exploited or manipulated by malicious actors. 
For instance, counterfactual explanations may be employed to identify adversarial samples with greater ease. 
Despite this, XAI systems remain susceptible to the biases inherent in the data and algorithms utilized, given that humans set the parameters for these systems~\cite{roselli2019managing}.
One potential solution is to involve multiple domain experts in lieu of a single expert.

\section*{Limitations of study}
The achievement of international harmonization in AI regulation represents a significant challenge, given the diverse regulatory approaches that exist across countries and regions. 
Nevertheless, there are indications of progress, including a growing consensus on high-risk applications and various international initiatives working towards harmonization.
The key challenges that must be addressed include the definition of AI, the balancing of innovation with risk mitigation, and the navigation of international tensions.
On a more positive note, there is an increasing political will for regulation, signs of trans-Atlantic convergence, and the potential influence of EU regulations globally.
Potential solutions include the establishment of global governance institutions, the adaptation of best practices from other high-tech sectors, and the development of open-source AI software. 
While full harmonization remains a challenging objective, there is a growing consensus towards more coordinated approaches to AI regulation through continued dialogue, cooperation, and flexible frameworks adaptable to this rapidly evolving technology. 
Our nested model is designed to facilitate international harmonization for AI regulations. 
The essential requirements identified by regulatory bodies in different countries may vary, but the underlying fundamental principle to satisfy these requirements remains consistent.

\section*{Conclusion}
In conclusion, our work is the first to bring together the disparate fields of AI and regulation. 
Our work establishes a common baseline for designing and validating AI under the umbrella of regulation, taking into account the complexity of the domain and the need for domain experts. 
Through this work, we are addressing discussions around the need to work at the intersection of AI and regulation to increase adoption in high-stakes domains such as healthcare.
Adopting the nested model for AI design and validation will help realize the full potential of today's state-of-the-art AI techniques in a complex, vast, but interconnected, and globalized world.   

\section*{Author Contributions}
Conceptualization, A.D., G.H.; Methodology. A.D., G.H.; Writing – Original Draft, A.D., Z.Y., and G.H.; Writing – Review \& Editing, A.D., Z.Y., and G.H.; Supervision, Z.Y. and G.H., Project Administration, G.H.

\section*{Declaration of Interests}
The authors declare no competing interests.

\printcredits
\pagebreak
\section*{STAR$\star$ METHODS}
\section*{Resource availability}
\subsection*{Lead contact}
Further information and requests should be directed to and will be fulfilled by the lead contact, Akshat Dubey (DubeyA@rki.de)
\subsection*{Materials availability}
Given the nature of the research, no datasets or codes are used or produced.
\subsection*{Data and code availability}
No data and code are involved in the research conducted.
\section*{Method details}
The research is focused on the evolving landscape of AI regulations. Although artificial intelligence (AI) has experienced significant growth in recent years, it has yet to achieve its full potential in real-world applications. This research is divided into four sections. The initial section examines the advantages of AI and elucidates the reasons behind the relatively slow rate of adoption of AI in critical and high-stakes domains. The advent of AI has brought with it a number of significant ethical and legal concerns. Consequently, there is a pressing need to address not only the regulatory policies that will facilitate the implementation of AI in real-world use cases but also how practitioners design and validate AI applications and workflows. The second part of the research is concerned with the potential of an audience-centric approach to AI, in conjunction with explainable artificial intelligence, to address significant concerns with AI regulations and satisfy the guidelines set out by regulatory authorities.  The third section of the research presents the XAI-QB, which is a question bank comprising algorithm-informed questions. These questions serve to assess and validate the AI, thereby fostering trust, faithfulness, and transparency in AI through the application of design principles.  In the fourth part, we have developed a nested model for AI design and validation on top of XAI-QB. The nested model contains five distinct layers: regulations, domain, data, model, and predictions. At each level, specific prototypical questions must be addressed, with due consideration given to the guidelines set forth by regulatory authorities. Additionally, illustrative examples have been provided.
\bibliographystyle{cas-model2-names}
\bibliography{cas-refs}
\end{document}